\documentclass[reprint, prd]{revtex4-2}
\usepackage{amsmath, amssymb}
\usepackage{graphicx,color,float,tabularx,siunitx}
\usepackage[caption=false]{subfig}
\usepackage{placeins}
\usepackage{rviewport}
\usepackage{orcidlink}
\definecolor{colorLink}{rgb}{0,0,180} 
\usepackage{hyperref}
\hypersetup{
   colorlinks = true,
   citecolor  = colorLink,
   urlcolor   = colorLink,
   linkcolor  = colorLink,
}

\DeclareSIUnit\year{yr}

\usepackage{float} 
\usepackage{graphicx}
\usepackage{subcaption}

\usepackage[sort&compress]{natbib}

\usepackage[shortcuts]{extdash}

\begin{document}
\title{Bifurcation and Critical Phenomena in Black Hole Thermodynamics
}

\author{Bidyut Hazarika \orcidlink{0009-0007-8817-1945}$^1$}

\email{rs\_bidyuthazarika@dibru.ac.in}

\author{Prabwal Phukon \orcidlink{0000-0002-4465-7974}$^1$,$^2$}
\email{prabwal@dibru.ac.in}

	\affiliation{$^1$Department of Physics, Dibrugarh University, Dibrugarh, Assam, 786004.\\$^2$Theoretical Physics Division, Centre for Atmospheric Studies, Dibrugarh University, Dibrugarh, Assam, 786004.\\}


\begin{abstract}
In this work, we treat black holes as bifurcation points and explore their thermodynamic phase structure using the framework of bifurcation theory which is a commonly used method from nonlinear dynamics. By constructing an appropriate bifurcating function, we analyze how black holes transition between different thermodynamic phases through changes in the number and stability of fixed points. Our study shows that stable fixed points correspond to thermodynamically stable black hole states, while unstable ones indicate instability and decay. The dynamical evolution of the system further supports this correspondence, with stable configurations approaching equilibrium and unstable ones diverging from it.

\end{abstract}


\maketitle                                                                      

\section{Introduction}
Black holes stand among the most intriguing consequences of Einstein’s General Theory of Relativity (GR), introduced in 1915. One of the most compelling confirmations of this theory came with the detection of gravitational waves observed by the LIGO experiment \cite{ligo}. This monumental discovery was soon complemented by the Event Horizon Telescope’s success in capturing direct images of supermassive black holes. These included the now-famous image of the black hole in galaxy M87 and the subsequent imaging of Sagittarius A* (SgrA*), the black hole residing at the core of our Milky Way \cite{m87a, m87b, m87c, m87d, m87e, m87f}. These observations, showing a dark shadow surrounded by a bright photon ring, offered deep insights into the geometry and nature of black holes and served as compelling tests of gravitational theories \cite{Shadow1,Shadow2,Shadow3,Shadow4}.

The development of black hole thermodynamics in the 1970s marked a major theoretical advancement, linking gravitational physics with the principles of thermodynamics \cite{Bekenstein:1973ur,Hawking:1974rv,Hawking:1975vcx,Bardeen:1973gs}. Since then, the field has grown considerably, with numerous theoretical contributions extending these foundational ideas \cite{Wald:1979zz,bekenstein1980black,Wald:1999vt,Carlip:2014pma,Wall:2018ydq,Candelas:1977zz,Mahapatra:2011si}. One especially compelling area of study involves the phase transitions of black holes \cite{Davies:1989ey,Hawking:1982dh,curir_rotating_1981,Curir1981,Pavon:1988in,Pavon:1991kh,OKaburaki,Cai:1996df,Cai:1998ep,Wei:2009zzf,Bhattacharya:2019awq,Kastor:2009wy,Dolan:2010ha,Dolan:2011xt,Dolan:2011jm,Dolan:2012jh,Kubiznak:2012wp,Kubiznak:2016qmn}. Various types of transitions have been explored, including the Davies transition \cite{Davies:1989ey}, the Hawking–Page transition between thermal anti–de Sitter space and black holes \cite{Hawking:1982dh}, transitions to extremal black hole states \cite{curir_rotating_1981,Curir1981,Pavon:1988in,Pavon:1991kh,OKaburaki,Cai:1996df,Cai:1998ep,Wei:2009zzf,Bhattacharya:2019awq}, and phenomena resembling Van der Waals phase transitions in the context of extended thermodynamic phase space \cite{Kastor:2009wy,Dolan:2010ha,Dolan:2011xt,Dolan:2011jm,Dolan:2012jh,Kubiznak:2012wp,Kubiznak:2016qmn}.  \\

Over the past two decades, black hole thermodynamics has undergone a significant evolution, especially with the realization that black holes in anti-de Sitter (AdS) space exhibit  strikingly analogous behaviour to some classical thermodynamics system. 
A milestone in this development was the identification of Van der Waals-type critical behavior in AdS black holes, where the cosmological constant is interpreted as a thermodynamic pressure and its conjugate as volume~\cite{Kubiznak:2012wp,Dolan:2010ha}. This perspective has led to the study of black hole chemistry~\cite{Kubiznak:2016qmn}. Phase transitions have also been studied from the restricted phase space thermodynamics and holographic thermodynamics perspectives~\cite{rp1,rp2,rp3,rp4,rp5,rp6,rp7,rp8,rp9,rp10,rp11}. Complementary to these macroscopic thermodynamic approaches, the thermodynamic geometry framework~\cite{Quevedo:2008ry,Akbar:2011qw,Hendi:2015cka,Sarkar:2006tg,Hendi:2015xya,Banerjee:2016nse,Bhattacharya:2017hfj,Bhattacharya:2019qxe,Gogoi:2023qni,Kumar:2012ve} has provided insights into the microscopic nature of black hole phase transitions~\cite{Wei:2015iwa}.
More exotic thermodynamic behaviors, including triple points and reentrant phase transitions, have also been reported in the context of rotating AdS black holes \cite{Altamirano:2013uqa}, while connections to condensed matter analogs have been explored through superfluid-like criticality \cite{Hennigar:2017apu}. 
The application of various aspects of nonlinear dynamics in the context of black hole physics has emerged as a fascinating area. The dynamical stability of black holes during transitions between different black hole phases was explored in Ref. \cite{nl1}, while the instabilities of thin black rings were investigated in Ref.\cite{nl2}. The onset of chaotic geodesic motion due to perturbations in black hole spacetimes, such as those introduced by surrounding discs or rings, was studied in Ref. \cite{nl3,nl4}. Thermal chaos in the context of black hole thermodynamics has also gained attention Ref. \cite{ch1,ch2,ch3,ch4}. 
In addition to these developments, black hole phase transitions have been examined from a topological perspective, which is closely connected to phase space dynamics Ref. \cite{t1,t2,t3}.

Our work builds on these foundations by introducing a bifurcation-based  perspective to study the emergence, stability, and criticality of black hole solutions. Bifurcation theory explores how the qualitative behavior of a dynamical system can change as parameters within the system are varied. Even in simple one-dimensional systems, where solutions typically converge to fixed points or diverge, interesting dynamics arise when a control parameter is introduced. As this parameter changes, fixed points can emerge or vanish, or their stability may shift phenomena collectively known as bifurcations. These transitions often correspond to critical thresholds in real-world systems. For instance, a vertical beam subjected to increasing weight may initially remain stable, but beyond a certain load, it loses stability and buckles.\cite{book}.  In this case, the weight serves as the control parameter, while the beam’s deflection reflects the system's state. Such examples illustrate how bifurcation analysis provides valuable insight into sudden changes and instabilities in both natural and engineered systems.\\

In this work, we investigate the thermodynamic phase structure of black holes using bifurcation theory.  To facilitate this analysis, we introduce a \textit{bifurcation function} defined as  
\begin{equation}
\dot{z} = \frac{d z}{d \tau} = -\frac{d M}{d z} + h \frac{d S}{d z},
\label{master}
\end{equation}  
where \( h > 0 \) is a control parameter we refer to as the \textit{bifurcation parameter}, \( M \) denotes the ADM mass, and \( S \) represents the black hole entropy. This functional form is inspired by the concept of off-shell free energy \cite{york1,york2}, with the parameter \( h \) playing a role analogous to the off-shell temperature, thereby justifying its positive definiteness. The variable \( z \) is a generalized measure of the black hole's size, which may correspond to quantities such as the event horizon radius or entropy. We also introduce an affine parameter \( \tau \), which serves to parameterize the evolution of the system and may be interpreted as an effective time variable in this dynamical framework.\\

 The central motivation of this work is to demonstrate that black hole phase transitions can be understood as natural manifestations of bifurcation phenomena much like phase transitions observed in condensed matter systems. Drawing on techniques from nonlinear dynamics, we propose a novel approach in which black holes are modeled as bifurcating systems, with their phase structure captured through bifurcation theory.
In this framework, we treat the black hole as a dynamical system characterized by a bifurcation function, and we introduce a control parameter \( h \), which plays the role of an order parameter analogous to external fields or temperature in conventional thermodynamic systems. Variations in thermodynamic variables specific to that system can lead to qualitative changes in the phase portrait.  These changes signify black hole phase transitions, which become evident in the shape of bifurcation diagrams. Moreover, this approach allows us to probe the stability of various black hole branches by analyzing the nature of the fixed points in the dynamical system. Stable fixed points correspond to thermodynamically stable black hole configurations, while unstable ones indicate branches that are dynamically or thermodynamically unfavorable.  We can use this scheme to classify black holes into different bifurcation class based on the number of fix points or shape of the bifurcation curve. 

\begin{figure*}[t!]
    \centering
    \subfloat[$h<h_c$\label{fig:subfig1}]{
        \includegraphics[width=0.3\textwidth]{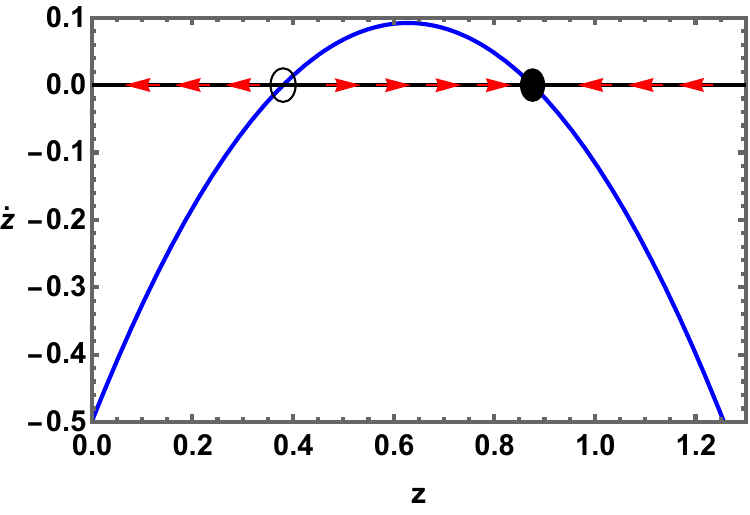}
    }\hfill
    \subfloat[$h=h_c$\label{fig:subfig2}]{
        \includegraphics[width=0.3\textwidth]{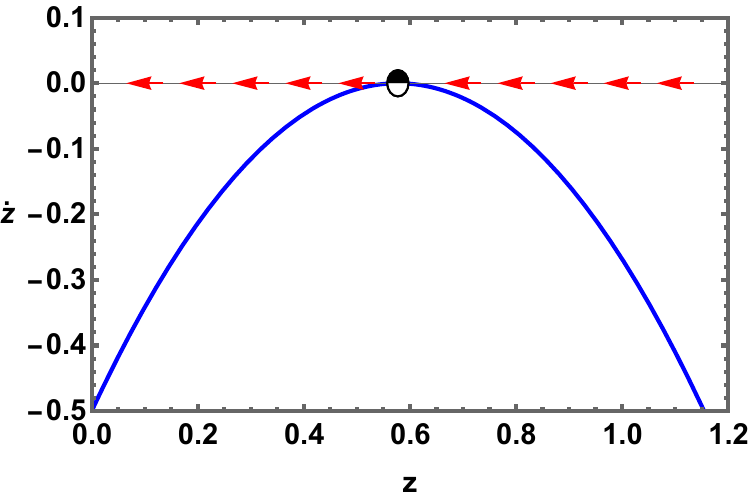}
    }\hfill
    \subfloat[$h>h_c$\label{fig:subfig3}]{
        \includegraphics[width=0.3\textwidth]{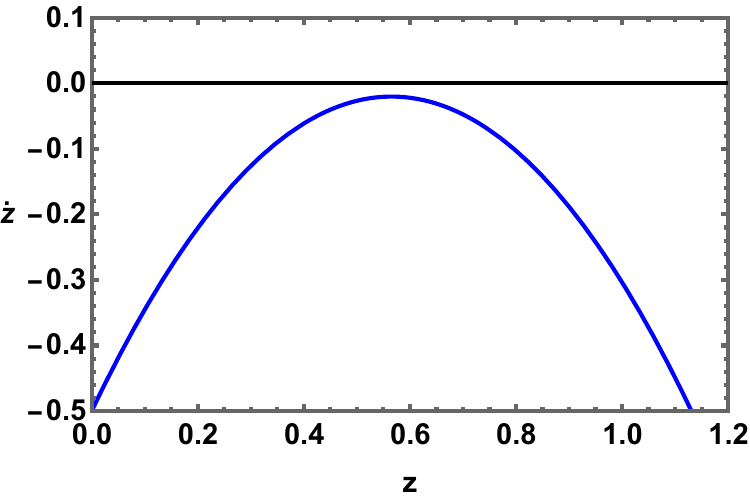}
    }
    \caption{Stability of the fix points}
    \label{fig:three_subfigs}
\end{figure*}

In the next section, we demonstrate how bifurcation theory applies to black hole thermodynamics. To illustrate this framework, we have selected four representative examples: Schwarzschild-AdS, RN-AdS, Euler–Heisenberg-AdS, and 6D Gauss-Bonnet-AdS black holes. These cases were chosen to highlight the versatility of our approach.  Each example exhibits a distinct bifurcation structure characterized by a different maximum number of fixed points in the associated dynamical system. For instance, the Schwarzschild-AdS black hole displays up to two fixed points, while the 6D Gauss-Bonnet-AdS black hole exhibits as many as five.
\section{Bifurcations in Black hole Thermodynamics}
\subsection{Schwarzschild AdS Black hole}
The line element of the Schwarzschild black hole in four dimensions is given by
\begin{multline}
ds^2 = -\left(1 - \frac{2M}{r} +\frac{r^2}{l^2}\right) dt^2 + \left(1 - \frac{2M}{r} +\frac{r^2}{l^2} \right)^{-1} dr^2 \\+ r^2 \left( d\theta^2 + \sin^2\theta\, d\phi^2 \right),
\label{eq:schwarzschild_metric}
\end{multline}
where \( M \) is the ADM mass of the black hole.

The mass \( M \) of the Schwarzschild black hole can also be expressed in terms of the event horizon radius \( r_+ \) as
\begin{equation}
M = \frac{r_+ \left(l^2+r_+^2\right)}{2 l^2}.
\label{m1}
\end{equation}
The rescaled mass $m$ with respect to $l$ is obtained as 
\begin{equation}
m= \frac{1}{2} z \left(z^2+1\right)
\label{m2}
\end{equation}
where $z=\frac{r_+}{l}$. 
Now using the bifurcating equation eq.  \ref{master}, we obtain
\begin{equation}
\dot{z}=-\frac{d m}{d z}+h \frac{d S}{d z}=- \frac{1}{2} \left(-4 \pi  h z+3 z^2+1\right)
\label{bi1}
\end{equation}
We have observed that the Schwarzschild black hole exhibits a saddle-node bifurcation, which represents a fundamental mechanism for the creation and annihilation of fixed points in dynamical systems\cite{book}. As the control parameter \( h \) is varied, two distinct fixed points gradually approach each other(Figure ~1(a)). The direction of the flow in the phase space depends on the sign of \(\dot{z}\): it moves to the right when \(\dot{z} > 0\) and to the left when \(\dot{z} < 0\). At points where \(\dot{z} = 0\), the system is at rest, and such locations are identified as fixed points. These fixed points can be classified based on the behavior of nearby trajectories. As illustrated in Fig.  1,  solid dots denote stable fixed points, where neighboring trajectories converge—these are often referred to as attractors or sinks. In contrast, open circles indicate unstable fixed points, from which nearby trajectories diverge—commonly known as repellers or sources.
When \( h = h_c \), the fixed points coalesce at a critical location \( z^* = 0 \), forming a single, half-stable fixed point. This scenario corresponds to the tangency of a parabola with the horizontal axis, as illustrated in Figure~1(b). For values \( h > h_c \), the system undergoes a topological change where the fixed points annihilate each other, and no real fixed points remain, as depicted in Figure~1(c). This delicate fixed point at the bifurcation threshold is structurally unstable, and its disappearance signals a qualitative shift in the system's dynamics.\\

The most common way to depict a bifurcation is by solving the equation \(\dot{z} = 0\) to obtain the fixed points \(z^*\) as a function of the control parameter \(h\). For \(h < h_c\), this yields two solutions for \(z^*\), corresponding to a stable and an unstable fixed point branch. Since \(h\) serves as the control (or bifurcation) parameter, it is conventionally plotted on the horizontal axis. The resulting plot of \(z^*\) versus \(h\) is referred to as the bifurcation diagram, and it is shown in Fig.  \label{2} for the case of a saddle-node bifurcation. To clearly distinguish the nature of the fixed points, stable branches are represented using solid lines, while unstable branches are shown using dashed lines.\\

\begin{figure}[h]
        \includegraphics[width=0.45\textwidth]{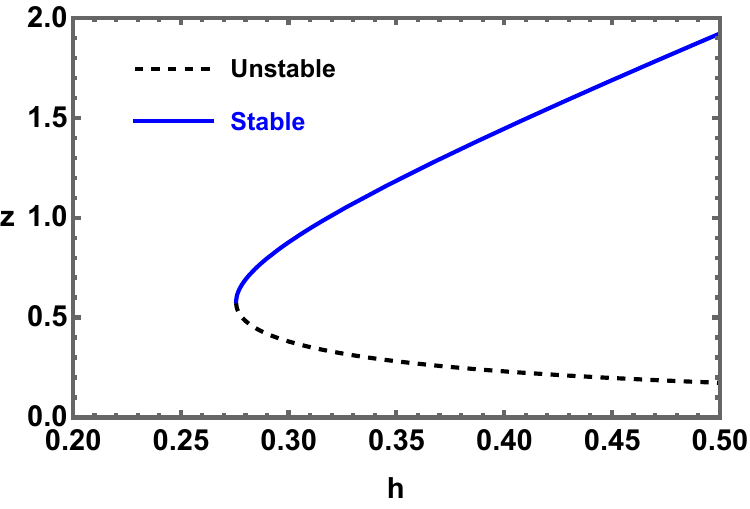}
        \caption{Bifurcation diagram for Schwarzchild black hole}
        \label{2}
\end{figure}
The terminology in bifurcation theory remains somewhat unsettled, with various authors using different names for the same type of bifurcation. For example, what is commonly known as a saddle-node bifurcation is also referred to as a \emph{fold bifurcation}, due to the characteristic shape of the bifurcation curve, or as a \emph{turning-point bifurcation}, highlighting the critical point where the direction of solution branches reverses. Although the term "saddle-node" is standard, it is not particularly descriptive in one-dimensional systems, as it originates from higher-dimensional phase spaces where saddle and node fixed points can interact and annihilate. In such contexts, the terminology reflects the nature of the colliding fixed points. Some literature even introduces more imaginative terms; for instance, Abraham and Shaw (1988) describe what is essentially a saddle-node bifurcation in reverse as a \emph{blue sky bifurcation}, emphasizing how fixed points can appear abruptly as a control parameter changes, as if emerging from nowhere.\\

\begin{figure}[h]
        \includegraphics[width=0.4\textwidth]{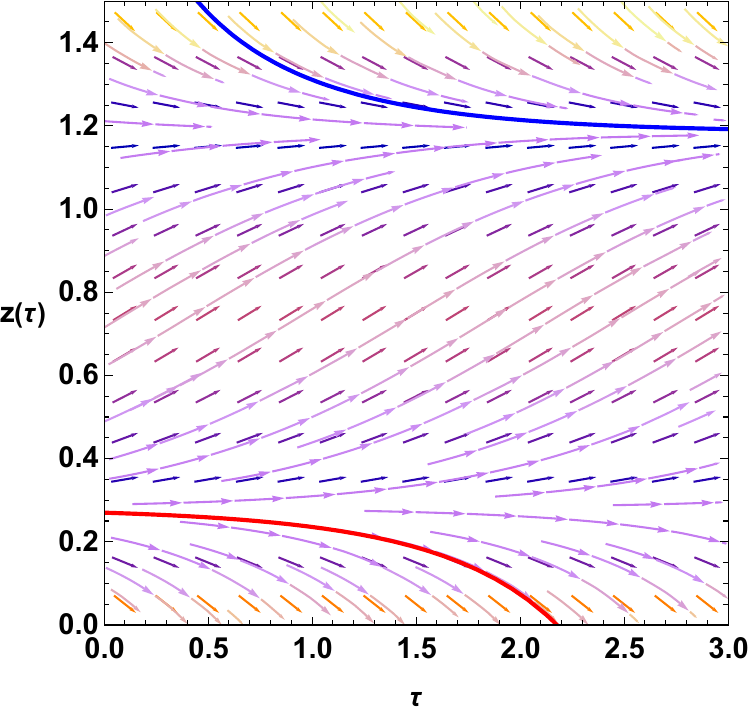}
        \caption{Dynamical analysis of Schwarzchild black hole}
        \label{slope}
\end{figure}

To analyze the dynamical behavior of the Schwarzschild black hole, we solve the differential equation eq. \ref{bi1}, numerically using the Runge–Kutta method.  For the control parameter value \(h = 0.35\), the system admits two fixed points located at \(z = 0.281362\) and \(z = 1.18471\). As an initial step, we construct the slope field of the system in the \((\tau, z)\) plane. In this representation, Eq.~\eqref{bi1} is interpreted such that each point \((\tau, z)\) defines a slope \(\mathrm{d}z/\mathrm{d}\tau\), corresponding to the tangent of a solution curve passing through that point. These slopes are visualized as line stream segments in Fig.~\ref{slope}.

\begin{figure*}[t!]
    \centering
    \subfloat[\label{fig:subfig1}]{
        \includegraphics[width=0.3\textwidth]{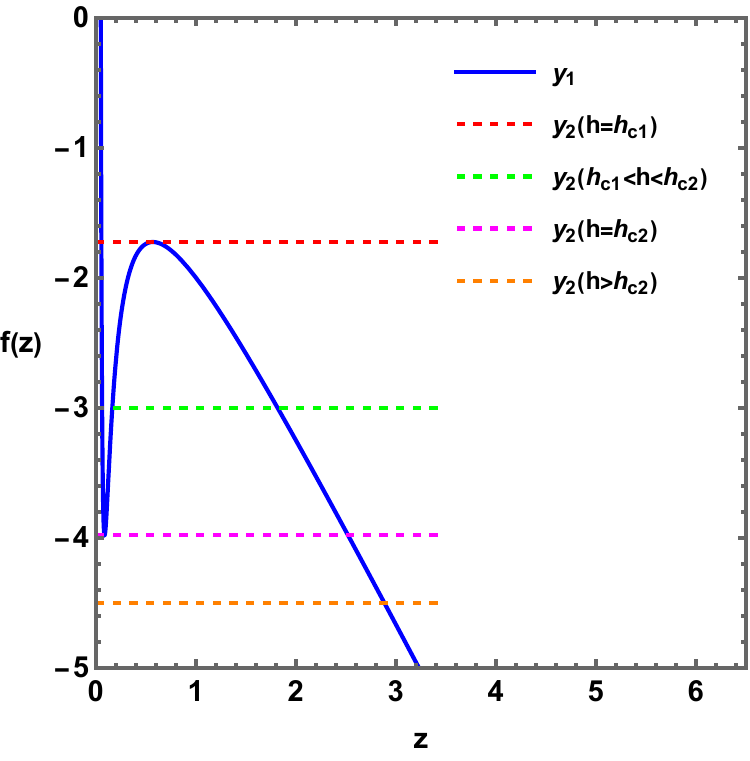}
    }\hfill
    \subfloat[\label{fig:subfig2}]{
        \includegraphics[width=0.3\textwidth]{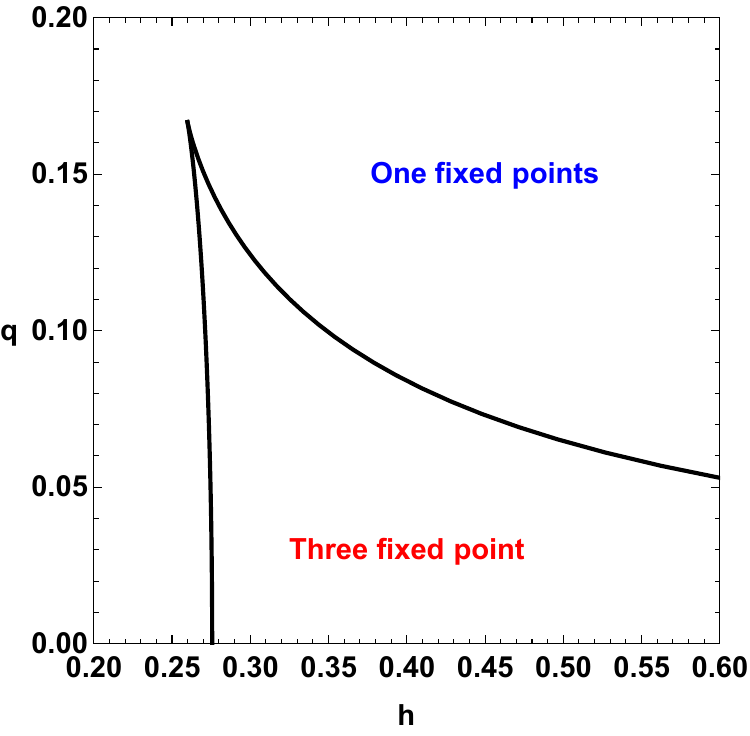}
    }\hfill
    \subfloat[\label{fig:subfig2}]{
        \includegraphics[width=0.3\textwidth]{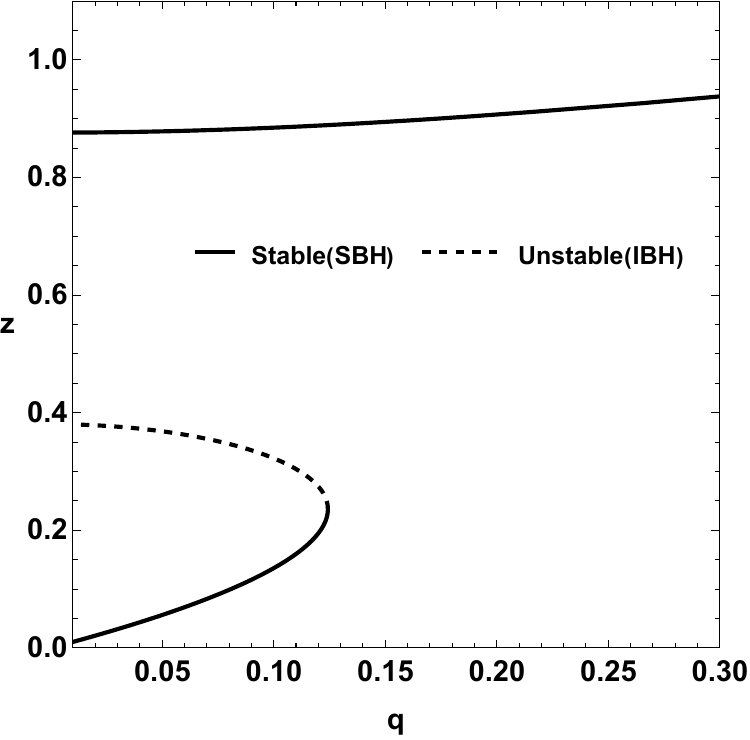}
    }\hfill
     \subfloat[$h_{c1}<h<h_{c2}$\label{fig:subfig1}]{
        \includegraphics[width=0.3\textwidth]{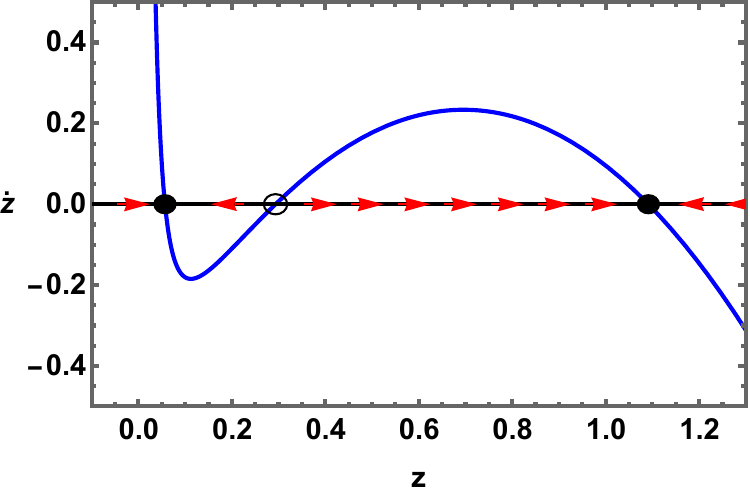}
    }\hfill
    \subfloat[$h=h_{c1}$\label{fig:subfig2}]{
        \includegraphics[width=0.3\textwidth]{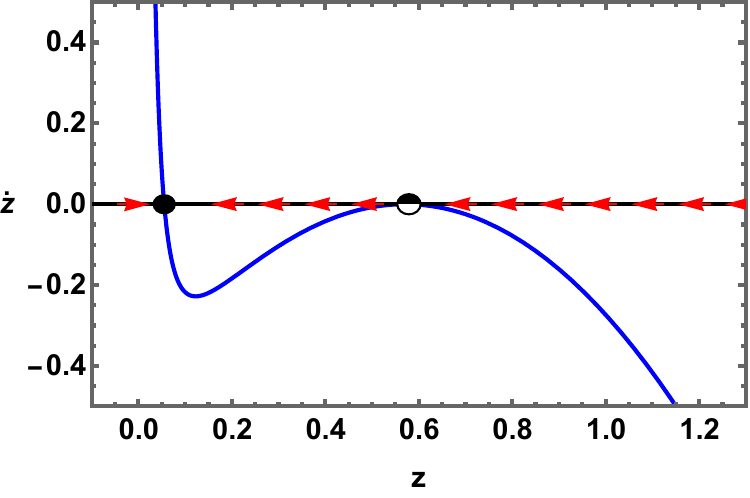}
    }\hfill
    \subfloat[$h>h_{c2}$\label{fig:subfig2}]{
        \includegraphics[width=0.3\textwidth]{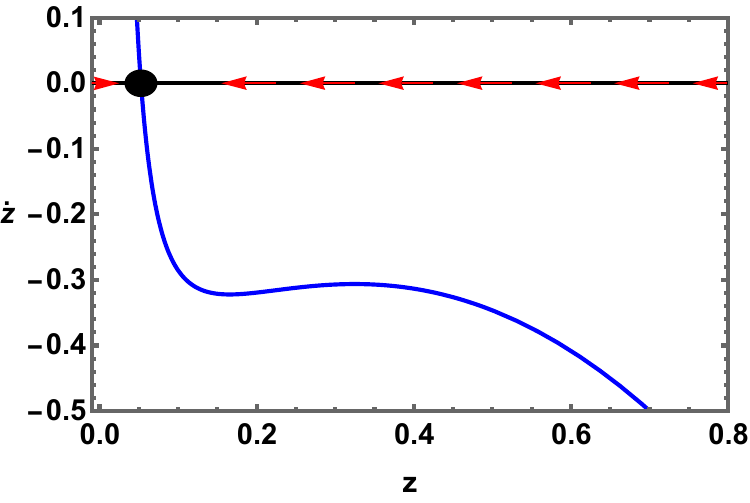}
    }\hfill
    \caption{Stability of the fix points of RN AdS black hole}
    \label{fig:three_subfigs}
\end{figure*}
To explore the evolution of the system, we select initial conditions \((z, \tau_0)\), guided by physical considerations—specifically, we require the black hole temperature to be positive at \(\tau = 0\), which constrains the initial choice of the radius. In Fig.~\ref{slope}, we present two representative solution trajectories. In particular, when the initial radius is chosen near the fixed point \(z = 1.18471\), the numerical integration yields the blue trajectory. It is evident from this plot that near the fixed point, the slope \(\mathrm{d}z/\mathrm{d}\tau\) approaches zero, indicating that the "velocity" of the system vanishes. This behavior signifies a state of equilibrium or stability. The blue curve asymptotically flattens out near \(z = 1.18471\), demonstrating that this point acts as a stable fixed point for the system described by Eq.~\eqref{bi1}.  In contrast, when the initial condition is chosen near the fixed point \(z = 0.281362\), the numerical integration results in the red trajectory shown in Fig.~\ref{slope}. As seen in the plot, the solution rapidly diverges from this fixed point; the slope \(\mathrm{d}z/\mathrm{d}\tau\) increases sharply, indicating a sudden change in the system's "velocity." This rapid deviation reflects the inherently unstable nature of the fixed point. Unlike the blue curve, which settles into a stable configuration, the red trajectory moves away from the vicinity of the fixed point, confirming that \(z = 0.281362\) acts as an unstable fixed point for the dynamical system governed by Eq.~\eqref{bi1}.\\

We now turn to a physical interpretation of the bifurcation structure observed in the Schwarzschild black hole system. Our analysis reveals that the system undergoes a saddle-node bifurcation, which gives rise to two distinct branches of fixed points: one stable and one unstable.  As we move from left to right along the \(z\)-axis in the phase diagram, each stable fixed point corresponds to a progressively larger black hole branch. For instance, in a scenario with four fixed points, the leftmost fixed point represents the smallest black hole, while the rightmost one corresponds to an ultra-large black hole configuration. By analyzing the direction of the flow arrows around each fixed point, we can directly infer the stability of the associated black hole branch. This provides an intuitive and efficient method for determining the thermodynamic stability of black hole solutions within the bifurcation framework. In this case, it can be easily observed that the stable fixed points correspond to the large black hole branch, while the unstable ones correspond to the small black hole branch. Therefore, the bifurcating behavior of the system effectively captures the essence of black hole phase transitions.  Furthermore, this approach can be linked to topological characteristics\cite{t2} if the flow arrows on either side of a fixed point point in the same direction, the winding number associated with that point is \(+1\); if they point in opposite directions, the winding number is \(-1\). It should be noted that the winding number associated with a fixed point directly corresponds to the winding number of the black hole branch  configuration represented by that point. \\
From a dynamical perspective, the higher fixed point, associated with the large black hole,  exhibits vanishing slope, indicating that the system settles into a stable configuration over time. This suggests that large Schwarzschild black holes evolve toward a stable equilibrium state. In contrast, the lower fixed poin representing a small black hole is dynamically unstable. As time progresses, the solution decays  from this point, indicating that small black holes tend to decay and lose stability with time.

\subsection{Bifurcations in RN AdS black hole}
\begin{figure*}[t!]
    \centering
    \subfloat[\label{rnslope1}]{
        \includegraphics[width=0.32\textwidth]{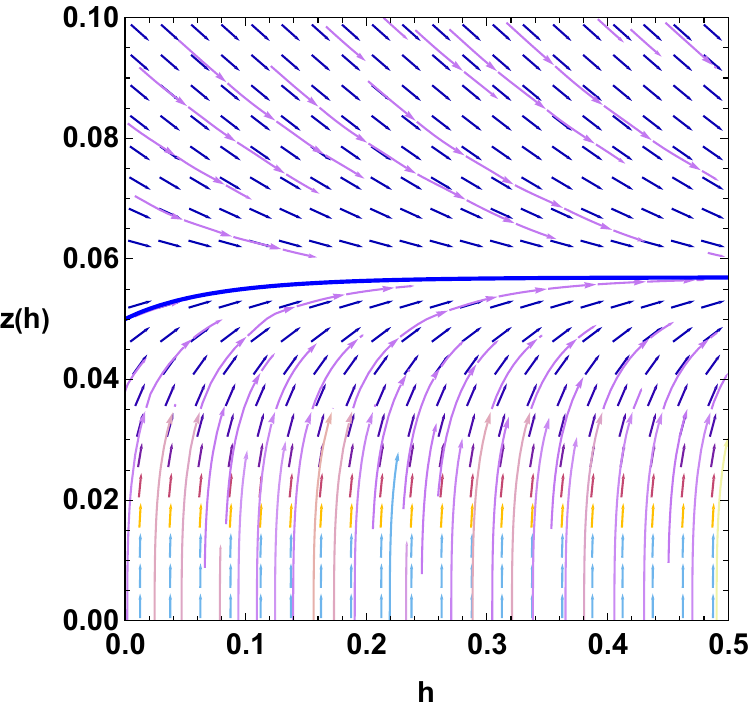}
    }\hfill
    \subfloat[\label{rnslope2}]{
        \includegraphics[width=0.32\textwidth]{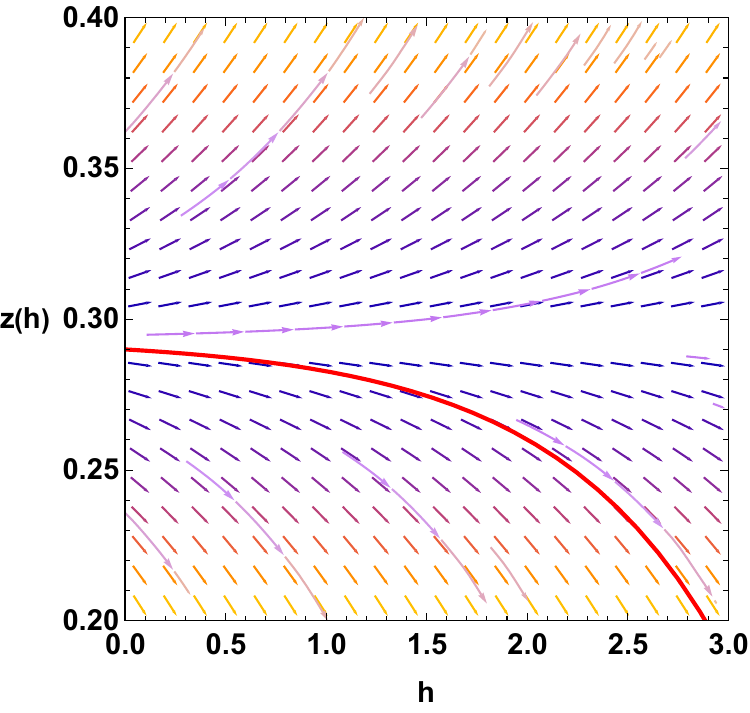}
    }\hfill
    \subfloat[\label{rnslope3}]{
        \includegraphics[width=0.32\textwidth]{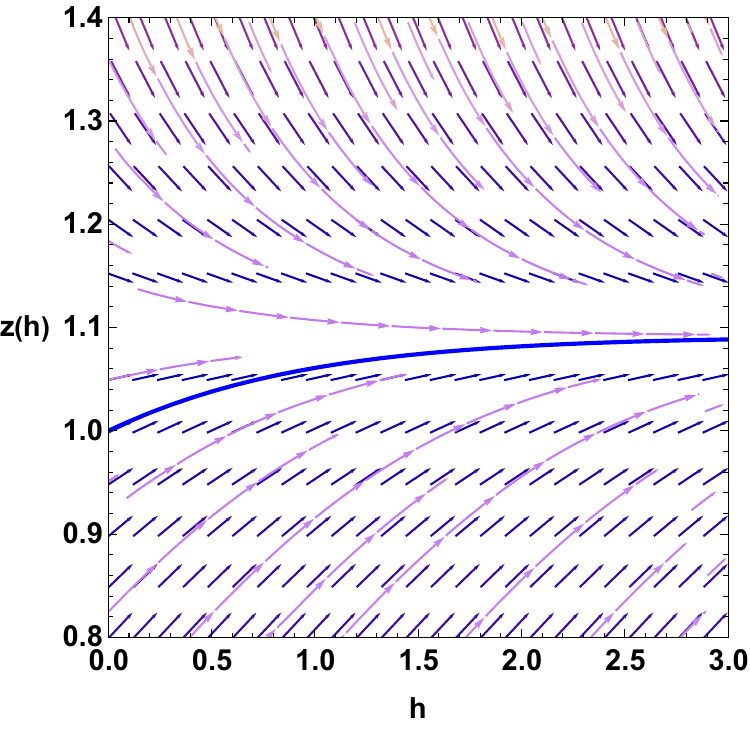}
    }\hfill
    \caption{Dynamical analysis  of RN AdS black hole}
    \label{fig:three_subfigs}
\end{figure*}
We now extend our bifurcation analysis to the Reissner–Nordström–Anti-de Sitter (RN-AdS) black hole. We consider a static, spherically symmetric spacetime described by the RN-AdS metric:
\begin{equation}
ds^2 = -f(r)\, dt^2 + \frac{dr^2}{f(r)} + r^2 \left( d\theta^2 + \sin^2\theta\, d\phi^2 \right),
\end{equation}
where the lapse function \( f(r) \) takes the form
\begin{equation}
f(r) = 1 - \frac{2M}{r} + \frac{Q^2}{r^2} + \frac{r^2}{l^2}.
\end{equation}
Here, \( M \) and \( Q \) represent the black hole's mass and electric charge, respectively, while \( l \) is the AdS radius. The mass parameter can be expressed in terms of the event horizon radius \( r_+ \) as
\begin{equation}
M = \frac{q^2}{2 z}+\frac{z^3}{2}+\frac{z}{2}
\end{equation}
where $q=Q/l$ and $z=r_+/l$

Using the mass expression,  the bifurcating equation for RN AdS black hole is written as 
\begin{equation}
\dot{z}=\frac{1}{2} \left(4 \pi  h z+\frac{q^2}{z^2}-3 z^2-1\right)
\end{equation}
or
\begin{equation}
\frac{\dot{z}}{z}=\frac{1}{2 z}\left(\frac{q^2}{z^2}-3 z^2-1\right)+2 h \pi
\label{bi2}
\end{equation}
To simplify the analysis, we fix the electric charge \( Q \) and examine how the system behaves under variations of the control parameter \( h \). Our goal is to determine the fixed points of Eq.~\eqref{bi2}. While these fixed points can, in principle, be obtained analytically using the general solution for the complex equations, the resulting expressions are cumbersome. Instead, a more transparent approach is to analyze the system graphically.

We consider the graphical intersection of the functions
\[
y_1(z) = \frac{1}{2z} \left(-1 + \frac{q^2}{z^2} - 3z^2\right), \quad \text{and} \quad y_2 = -2\pi h,
\]
plotted on the same axes. The fixed points of Eq.~\eqref{bi2} correspond to the points where these two curves intersect. Depending on the value of \( h \), the number of intersections may vary: the system can exhibit one or three fixed points, as illustrated in Fig. 4.

The critical scenario arises when the horizontal line \( y_2 \) becomes tangent to the curve \( y_1(z) \), signaling the occurrence of a saddle-node bifurcation. This happens when the line touches the local extremum of \( y_1(z) \). The local maximum of \( y_1(z) \) occurs at
\[
z_c = \frac{\sqrt{1 + \sqrt{1 - 36q^2}}}{\sqrt{6}},
\]
and the corresponding critical value of \( y_1 \) is
\[
y_{1,\text{max}} = -\frac{\sqrt{6} \left(1 - 12q^2 + \sqrt{1 - 36Q^2}\right)}{\left(1 + \sqrt{1 - 36q^2}\right)^{3/2}}.
\]
Thus, the saddle-node bifurcation takes place at the critical value \( h = h_c \), where \( h_c \) is determined by the condition \( y_2 = y_{1,\text{max}} \). This defines the threshold at which two fixed points coalesce and annihilate, marking the onset of a qualitative change in the system's dynamics.Hence $h_c$ expressions are
\begin{equation}
h_{c1}=\frac{\sqrt{6} \left(-12 q^2+\sqrt{1-36 q^2}+1\right)}{2 \pi  \left(\sqrt{1-36 q^2}+1\right)^{3/2}}
\label{hc1}
\end{equation}

\begin{equation}
h_{c2}=\frac{\sqrt{6} \left(12 q^2+\sqrt{1-36 q^2}-1\right)}{\left(1-\sqrt{1-36 q^2}\right)^{3/2}}
\label{hc1}
\end{equation}

As illustrated in Fig.~4(a), the region enclosed between $h_{c_1},h_{c2}$ contains three fixed points, whereas outside this region, the system admits only a single fixed point.
We now examine the bifurcation structure in the \((q, h)\) parameter space by plotting the critical curves \(h = h_{c_1}(q)\) and \(h = h_{c_2}(q)\) in Fig. 4(b). These curves delineate the boundaries where saddle-node bifurcations occur. Notably, both curves converge tangentially at the point \((q, h) = (0.260, 0.1667)\), a distinguished location referred to as a \emph{cusp point}.  The regions enclosed by these curves correspond to different dynamical regimes, each characterized by a distinct number of fixed points. Along the bifurcation curves (excluding the cusp), the system undergoes saddle-node bifurcations, indicating a transition between one and three fixed points. However, the cusp point represents a more intricate phenomenon,  a codimension-2 bifurcation, where two control parameters, \(q\) and \(h\), must be finely tuned simultaneously to reach this bifurcation condition. In contrast to the simpler codimension-1 bifurcations in the earlier case(which depend on varying a single parameter), codimension-2 bifurcations signify a higher degree of structural complexity in the phase space of the system.
In Fig.~4(c), we present the stability diagram by plotting the bifurcation profile in the \(z\)–\(q\) plane for a fixed value of \(h\). The resulting structure resembles a broken pitchfork, where the bifurcation curve splits into distinct segments. The upper segment comprises entirely stable fixed points, indicating a smooth and continuous evolution of the system as \(q\) increases. In contrast, the lower segment features both stable and unstable branches, but the stable portion of this lower segment cannot be accessed through small perturbations; it requires a significant deviation from equilibrium to reach.
Physically, the fixed points with larger \(z\) values correspond to large black hole (LBH) phases, which exhibit thermodynamic stability. The intermediate black hole (IBH) phase associated with the middle branch is found to be unstable, while the small black hole (SBH) branch,  located at the lowest \(z\) values is  stable. 

\begin{figure}[h]
        \includegraphics[width=0.45\textwidth]{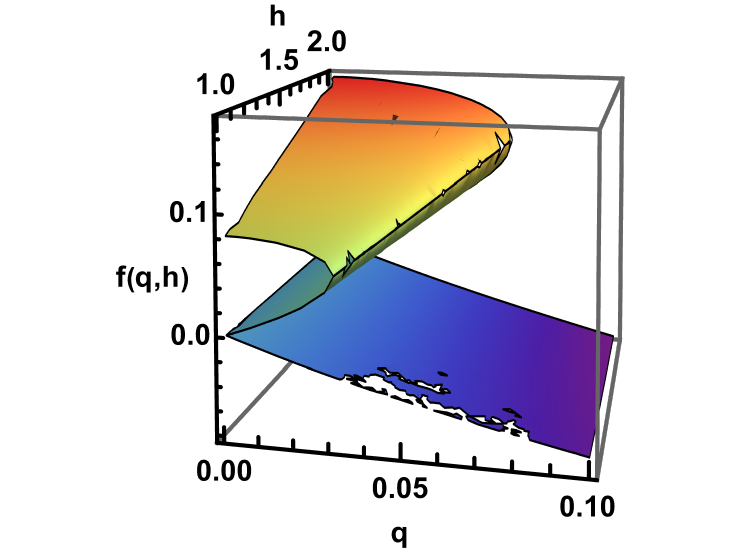}
        \caption{Cusp catastrophe surface}
        \label{3d1}
\end{figure}

Another compelling approach to represent the bifurcation behavior is through a three-dimensional visualization called as cusp catastrophe surface as shown in Fig.\ref{3d1}.  By plotting the fixed points \( z^* \) as a function over the parameter space \((q, h)\), one obtains a surface that reveals the full structure of the system’s dynamical responses. This surface, often referred to in dynamical systems theory as a cusp catastrophe surface, illustrates how the system transitions between different regimes of stability. The folds of this surface,  where multiple fixed points merge or vanish project down to the familiar bifurcation curves in the \((q, h)\) plane, serving as the boundaries between distinct dynamical phases. Cross-sectional slices of this surface provide additional insights: fixing \( h \) yields a profile in the \((z, q)\) plane, while fixing \( q \) gives a diagram in the \((z, h)\) plane. These sections correspond to the various bifurcation diagrams discussed previously, but now they are unified within a higher-dimensional geometric framework.

To examine the dynamical evolution of the RN-AdS black hole system, we solve the equation \eqref{bi2} numerically using suitable initial conditions \((z, \tau_0)\), guided by physical requirements. For a fixed value of the control parameter \(h\), our analysis reveals the presence of three distinct fixed points in the \(z\)-space, corresponding to three branches of black holes. Among these, the smallest and largest values of \(z\) are identified as stable fixed points representing the small black hole (SBH) and large black hole (LBH) brancheswhile the intermediate fixed point corresponds to an unstable configuration, associated with the intermediate black hole (IBH) branch.

The dynamical behavior of the system near these fixed points mirrors that of the Schwarzschild black hole scenario discussed earlier. When the initial radius is selected close to the LBH fixed point, the numerical integration yields a trajectory that quickly settles into a stable configuration, as the slope \(\mathrm{d}z/\mathrm{d}\tau\) vanishes near the fixed point. This confirms the attractor nature of the LBH state. A similar stabilization behavior is observed for initial conditions near the SBH fixed point, indicating its dynamical stability under small perturbations. In contrast, when the system is initialized near the IBH fixed point, the solution rapidly diverges, with \(\mathrm{d}z/\mathrm{d}\tau\) increasing sharply, reflecting a loss of stability. This is evidenced by the red trajectory in Fig.~\ref{rnslope2}, which moves away from the vicinity of the IBH point, verifying its role as a repeller. These results, consistent with the qualitative features of saddle-node bifurcations.

\subsection{Bifurcations  in Euler Heisenberg AdS black hole}
\begin{figure*}[t!]
    \centering
    \subfloat[$h_{c2}<h<h_{c3}$\label{egh1}]{
        \includegraphics[width=0.45\textwidth]{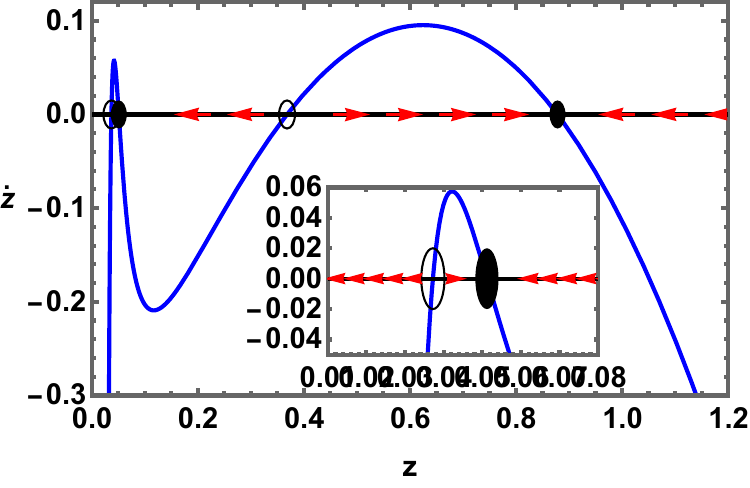}
    }\hfill
    \subfloat[$h=h_{c2}$\label{eh2}]{
        \includegraphics[width=0.45\textwidth]{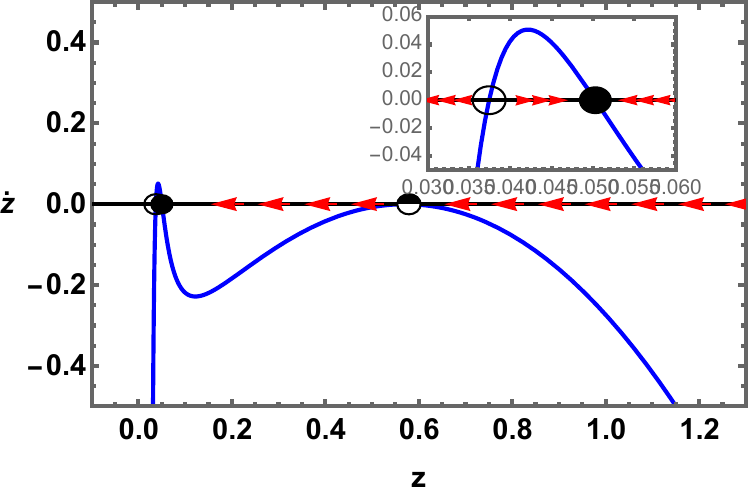}
    }\hfill
    \subfloat[$h=h_{c1}$\label{fig:subfig2}]{
        \includegraphics[width=0.45\textwidth]{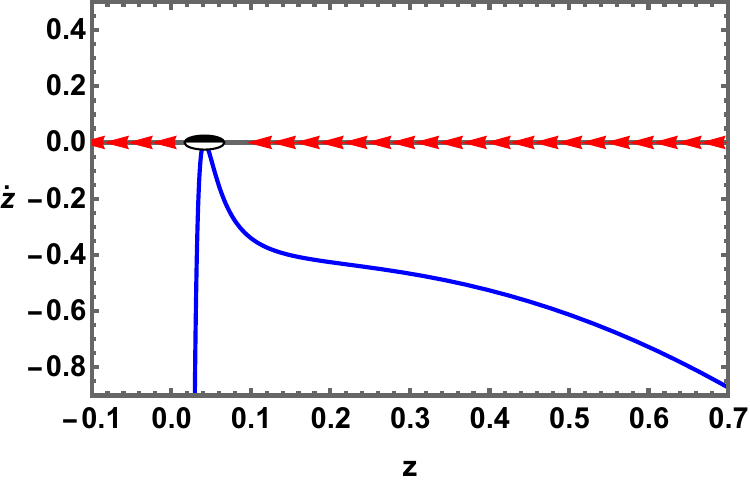}
    }\hfill
     \subfloat[$h_{c1}<h<h_{c2}$\label{fig:subfig1}]{
        \includegraphics[width=0.45\textwidth]{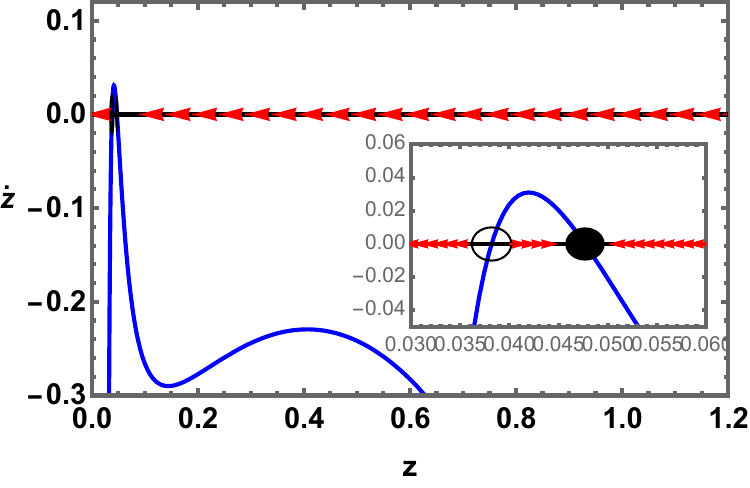}
    }\hfill
   \caption{Stability of the fixed points for the EH-AdS black hole with \(\tilde{\alpha} = 0.0016\). Solid dots indicate stable fixed points, while transparent dots represent unstable ones. The number of fixed points changes as the parameter \(h\) varies, with three critical values identified: \(h_{c1} = 0.08202\), \(h_{c2} = 0.27461\), and \(h_{c3} = 0.6388\). }
    \label{fig:three_subfigs}
\end{figure*}
The four-dimensional Euler-Heisenberg-AdS (EHAdS) black hole is a static, spherically symmetric solution to the Einstein equations with a nonlinear electromagnetic field described by the Euler-Heisenberg Lagrangian~\cite{Salazar:1987ap,Magos:2020ykt}. This nonlinear modification to Maxwell's theory arises from quantum electrodynamics (QED) corrections and is characterized by the Euler-Heisenberg (EH) parameter \( a \), which is given by
\begin{equation}
a = \frac{8\alpha^2}{45 m^4},
\end{equation}
where \(\alpha\) is the fine-structure constant and \(m\) is the electron mass, with natural units \(c = \hbar = 1\).

The spacetime metric is expressed as
\begin{equation}
ds^2 = -f(r)\, dt^2 + \frac{dr^2}{f(r)} + r^2\, d\Omega^2,
\end{equation}
where the lapse function \(f(r)\) takes the form
\begin{equation}
f(r) = 1 - \frac{2M}{r} + \frac{Q^2}{r^2} + \frac{\Lambda r^2}{l^2} - \frac{\alpha Q^4}{20 r^6}.
\end{equation}
Here, \(M\) is the black hole mass, \(Q\) is the electric charge.

The mass of the black hole, in terms of the event horizon radius \(r_+\), is given by
\begin{equation}
M = \frac{-\alpha  l^2 q^4+20 l^2 Q^2 r_+^4+20 l^2 r_+^6+20 r_+^8}{40 l^2 r_+^5}
\end{equation}
This expression reduces to the familiar Reissner-Nordström-AdS case in the limit \(\alpha \to 0\), and incorporates the leading-order QED correction from the Euler-Heisenberg theory. The rescaled bifurcating equation is calculated as :
\begin{equation}
\frac{d z}{d \tau}=\frac{1}{8} \left(4 \left(4 \pi  h z+\frac{q^2}{z^2}-3 z^2-1\right)-\frac{q^4 \tilde{\alpha }}{z^6}\right)
\label{bieh}
\end{equation}
where $q=\frac{Q}{l},  z=\frac{r_+}{l},  \tilde{\alpha}=\frac{\alpha}{l}$.
\begin{figure}[!t]
\centering
\includegraphics[width=\linewidth]{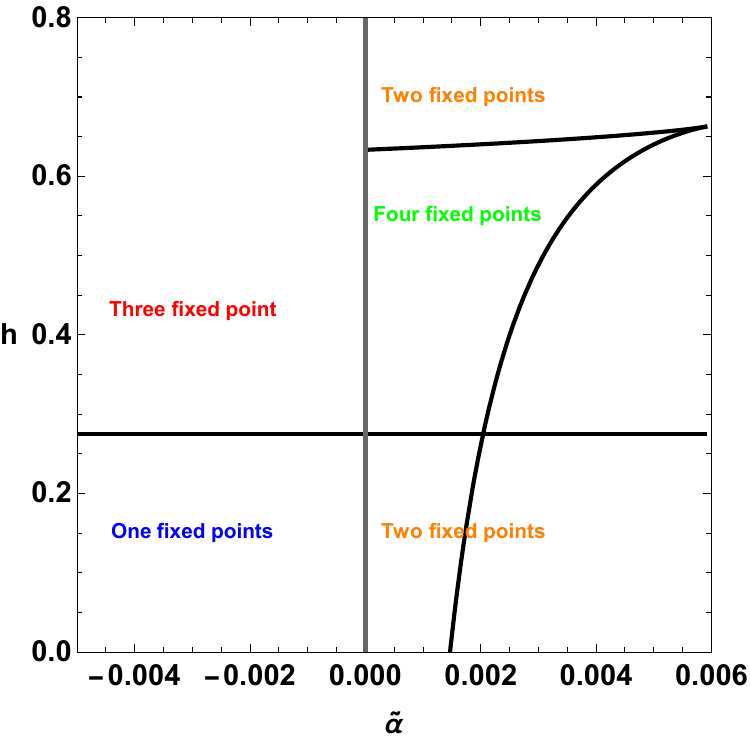}  
\caption{Bifurcation diagram : Number of fix points.  We have kept q=0.05 fixed.}
\vspace{0.2cm}
\includegraphics[width=\linewidth]{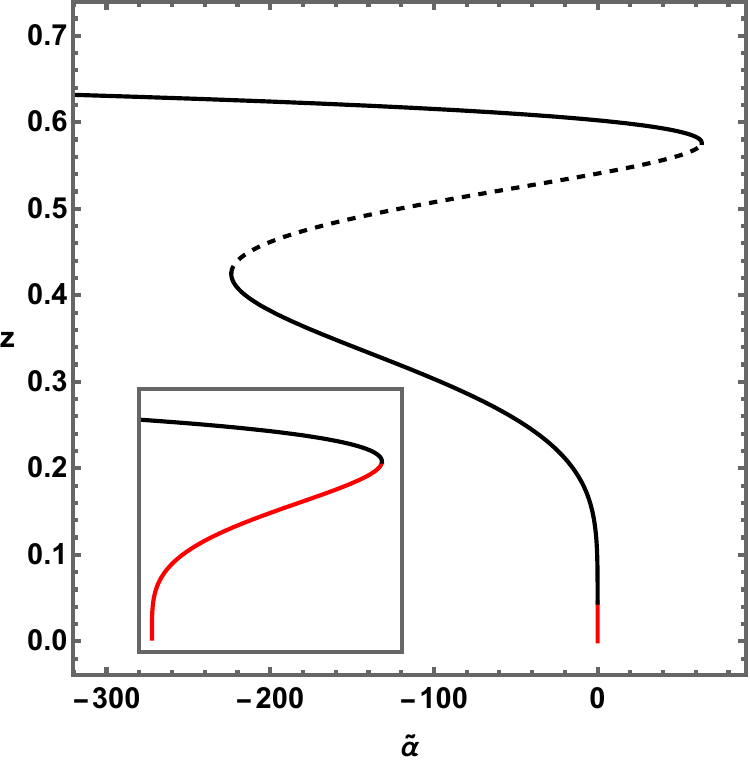}  

\caption{Bifurcation diagram : Stability of fix points for positive value of $\tilde{\alpha}$.  We have kept q=0.05 and h=0.275 fixed.}
\label{fig0}
\end{figure}

In the case of the EH black hole,  the bifurcation structure exhibits imperfect bifurcation behavior. Depending on the value of the EH parameter \(\tilde{\alpha}\) and bifurcating parameter h, the system admits different number of fixed points.  For positive value of \(\tilde{\alpha}\),  depending upon h value there are either four or two fixed points. Interestingly, for positive values of \(\tilde{\alpha}\), the system exhibits three \emph{half-stable} fixed points. Two of these appears during the transition from the four fixed point region to the two fixed point region, and the other emerges in the transition from the two fixed point region to the regime with no fixed points. The stability of these fixed points is illustrated in Fig.7, where the behavior of the system near each fixed point can be clearly observed.

In Fig.8,  we depict the regions in the \((h, \tilde{\alpha})\) parameter space where the number of fixed points changes.  For positive values of \(\tilde{\alpha}\), there exists a narrow region that supports four fixed points. Outside this domain, the system supports either two or no fixed points. The bifurcation diagram is shown in Fig.9. In this plot, black solid curves correspond to stable fixed point branches, while red solid lines and dashed black curves represent unstable branches. Based on this stability structure, we observe that the intermediate black hole (IBH) and ultra-large black hole (ULBH) branches are thermodynamically stable, while the small black hole (SBH) and large black hole (LBH) branches are thermodynamically unstable.

On the other hand, for negative values of \(\tilde{\alpha}\), the bifurcation diagram closely resembles the one observed for the RN-AdS black hole. As shown in Fig.8 the system exhibits either one or three fixed points. The corresponding bifurcation diagram takes the form of a broken pitchfork, similar to the RN-AdS case, with two stable branches and one unstable branch.  Furthermore, a dynamical analysis similar to that in the RN-AdS case can be performed. The evolution of the system confirms that the dynamical stability behavior is consistent with the thermodynamic stability inferred from the fixed point structure.

\subsection{Bifurcations in 6 D Gauss Bonnet AdS black hole}
\begin{figure*}[t!]
    \centering
    \subfloat[$h_{c2}<h<h_{c3}$\label{egh1}]{
        \includegraphics[width=0.3\textwidth]{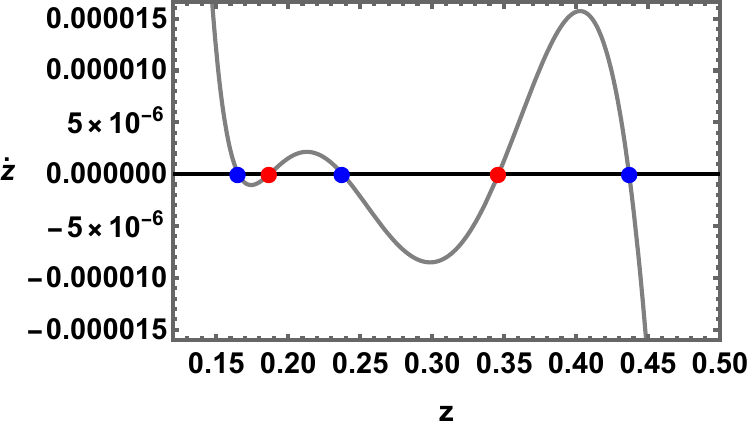}
    }\hfill
    \subfloat[$h_{c3}<h<h_{c4}$\label{eh2}]{
        \includegraphics[width=0.3\textwidth]{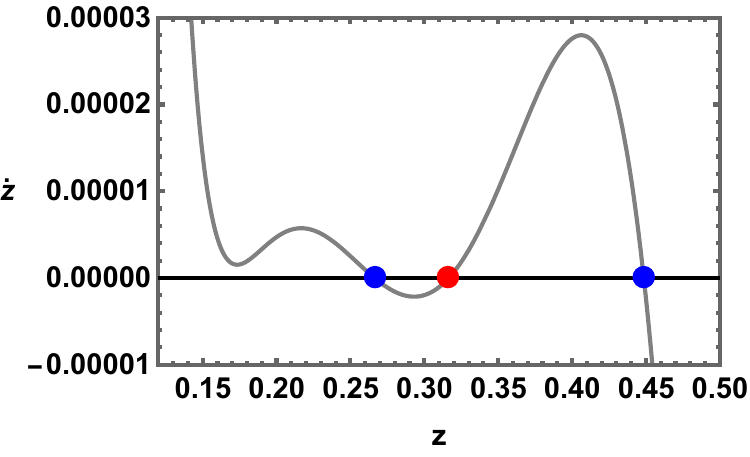}
    }\hfill
    \subfloat[$h=h_{c1}$\label{fig:subfig2}]{
        \includegraphics[width=0.3\textwidth]{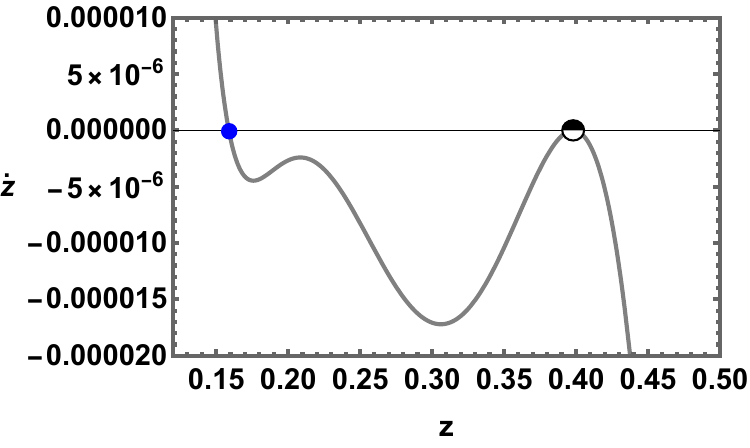}
    }\hfill
     \subfloat[$h=h_{c2}$\label{fig:subfig1}]{
        \includegraphics[width=0.3\textwidth]{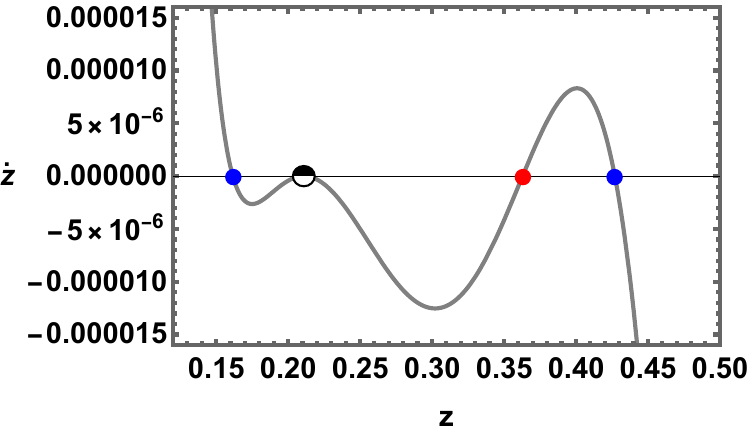}
    }\hfill
     \subfloat[$h=h_{c3}$\label{fig:subfig2}]{
        \includegraphics[width=0.3\textwidth]{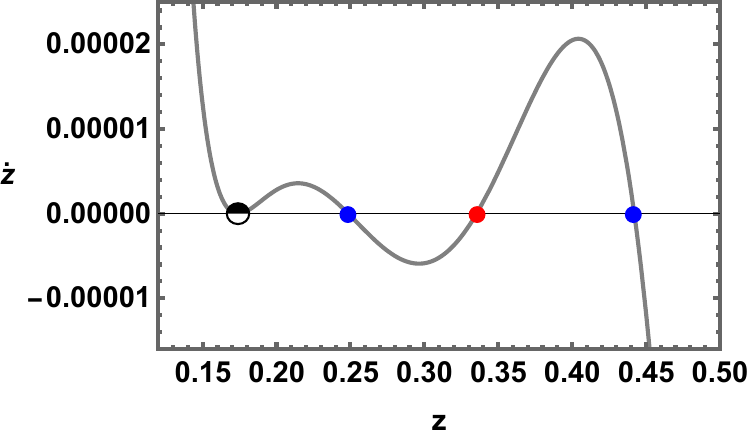}
    }\hfill
     \subfloat[$h=h_{c4}$\label{fig:subfig1}]{
        \includegraphics[width=0.3\textwidth]{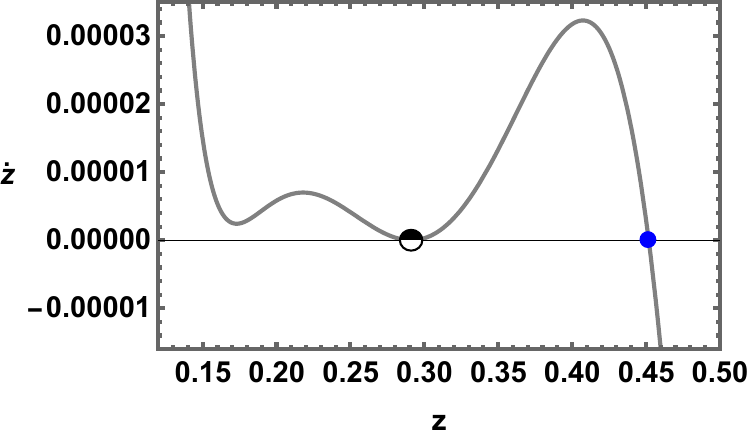}
    }\hfill
   \caption{Stability of the fixed points for the EH-AdS black hole with \(\tilde{\alpha} = 0.00815\) and  \(q = 0.00205\). Blue dots indicate stable fixed points, while red dots represent unstable ones. The number of fixed points changes as the parameter \(h\) varies, with four critical values identified: \(h_{c1} = 0.380148\), \(h_{c2} = 0.380229\),  \(h_{c3} = 0.380347\) and  \(h_{c4} =0.380456\)}
    \label{fig:three_subfigs}
\end{figure*}
We consider Gauss-Bonnet-AdS black hole in a $D$-dimensional spacetime, described by the metric~\cite{Cai:2013qga}
\begin{align}
\label{line_element_dD}
ds^2 = -f(r)\, dt^2 + \frac{dr^2}{f(r)} + r^2 h_{ij} dx^i dx^j,
\end{align}
where $h_{ij}$ is the metric of a symmetric space with constant curvature $(D-2)(D-3)k$, and $k=1, 0, -1$ corresponds to spherical, flat, and hyperbolic topologies, respectively. Focusing on the spherical case ($k=1$), the metric function takes the form
\begin{widetext}
\begin{equation}
f(r) = 1 + \frac{r^2}{2 \tilde{\alpha}} \biggl(1 -
 \sqrt{1 + \frac{64\pi M \tilde{\alpha}}{(D-2) r^{D-1}} -\frac{2 q^2 \tilde{\alpha}}{(D-2)(D-3) r^{2D-4}} + \frac{8 \tilde{\alpha} \Lambda}{(D-1)(D-2)}} \biggr)
\end{equation}
\end{widetext}
where $\tilde{\alpha} = (D-3)(D-4)\alpha$, with $\alpha$ the Gauss-Bonnet coupling constant, $M$ the black hole mass, and $q$ the electric charge. To ensure a real-valued metric, the expression under the square root must remain positive, which imposes an upper bound on $\alpha$, denoted as $\alpha_M$.

The event horizon radius $r_+$ is determined from the condition $f(r_+)=0$, yielding the mass:
\begin{multline}
M = \frac{r_+^{-D-5}}{16\pi (D-2) l^2} \biggl((D-3)(D-2) r_+^{2D} \\ \left[(\alpha(D-4)(D-3) + r_+^2) l^2 + r_+^4 \right] + 2 l^2 q^2 r_+^8 \biggr).
\end{multline}

The corresponding Bekenstein-Hawking entropy is
\begin{align}
S &= \frac{r_+^{D-2}}{4} \left[1 + \frac{2\alpha (D-2)(D-3)}{r_+^2} \right].
\end{align}
The rescaled bifurcating equation is 
\begin{equation}
\dot{z}=\frac{12 z^4 \tilde{\alpha } (32 \pi  h z-3)-2 z^6 \left(-16 \pi  h z+15 z^2+9\right)+3 q^2}{32 \pi  z^4}
\end{equation}
In this case, we observe a highly intricate bifurcation structure, with the system admitting up to five fixed points. The number of fixed points depends sensitively on both the Gauss-Bonnet (GB) coupling parameter \(\tilde{\alpha}\) and the bifurcation parameter \(h\). For specific values of \(\tilde{\alpha}\), the number of fixed points varies with \(h\), transitioning between one, three, and five fixed points.

Remarkably, the system exhibits four \emph{half-stable} fixed points. Two of these appear during the transition from a region with one fixed point to a region with three, while the other two emerge during the transition from three to five fixed points. For instance, by fixing \(q = 0.00205\) and \(\tilde{\alpha} = 0.00815\), we identify four half-stable fixed points at the following critical values of \(h\): 
\[
h_{c1} = 0.380148, \quad h_{c2} = 0.380229,
\]
\[  h_{c3} = 0.380347, \quad h_{c4} = 0.380456.\]
The number of fixed points in different regimes of \(h\) is summarized as follows:
\begin{itemize}
  \item For \(h < h_{c1}\) and \(h > h_{c4}\), the system admits one fixed point.
  \item For \(h_{c1} < h < h_{c2}\) and \(h_{c3} < h < h_{c4}\), there are three fixed points.
  \item In the intermediate region \(h_{c2} < h < h_{c3}\), the system admits five fixed points.
\end{itemize}

The stability properties of these fixed points are illustrated in Fig.~10, where the local behavior of the system near each fixed point is explicitly shown.

In Fig.~11 a, we present the regions in the \((h, \tilde{\alpha})\) parameter space where the number of fixed points changes. The four colored curves correspond to the four positive solutions of the equation \(\dot{z} = 0\). As evident from the figure, the bifurcation structure is complex and rich. 

The global bifurcation diagram is depicted in Fig.~11 b In this plot, solid black curves represent branches of stable fixed points, while dashed black curves correspond to unstable ones. From the stability structure, it is evident that the intermediate black hole (IBH) and ultra-large black hole (ULBH) branches are thermodynamically unstable, whereas the remaining branches correspond to thermodynamically stable configurations.

Here also a dynamical analysis similar to previous cases can be performed. The evolution of the system confirms that the dynamical stability behavior is consistent with the thermodynamic stability inferred from the fixed point structure.

\begin{figure*}[t!]

    \subfloat[Bifurcation diagram : Number of fix points.  We have kept q=0.05  fixed.\label{egh1}]{
        \includegraphics[width=0.4\textwidth]{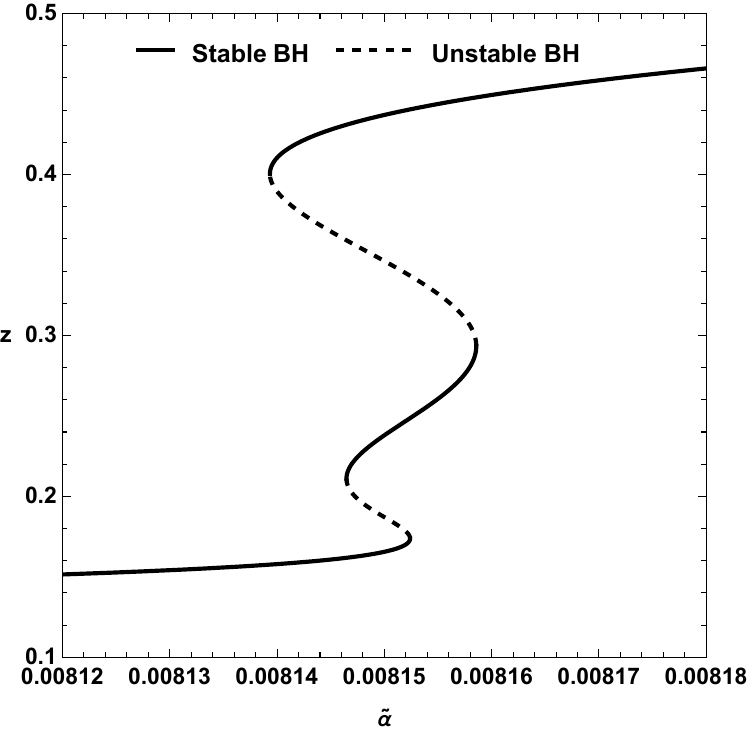}
    }
    \subfloat[Bifurcation diagram : Stability of fix points.  We have kept q=0.05  fixed.\label{eh2}]{
        \includegraphics[width=0.4\textwidth]{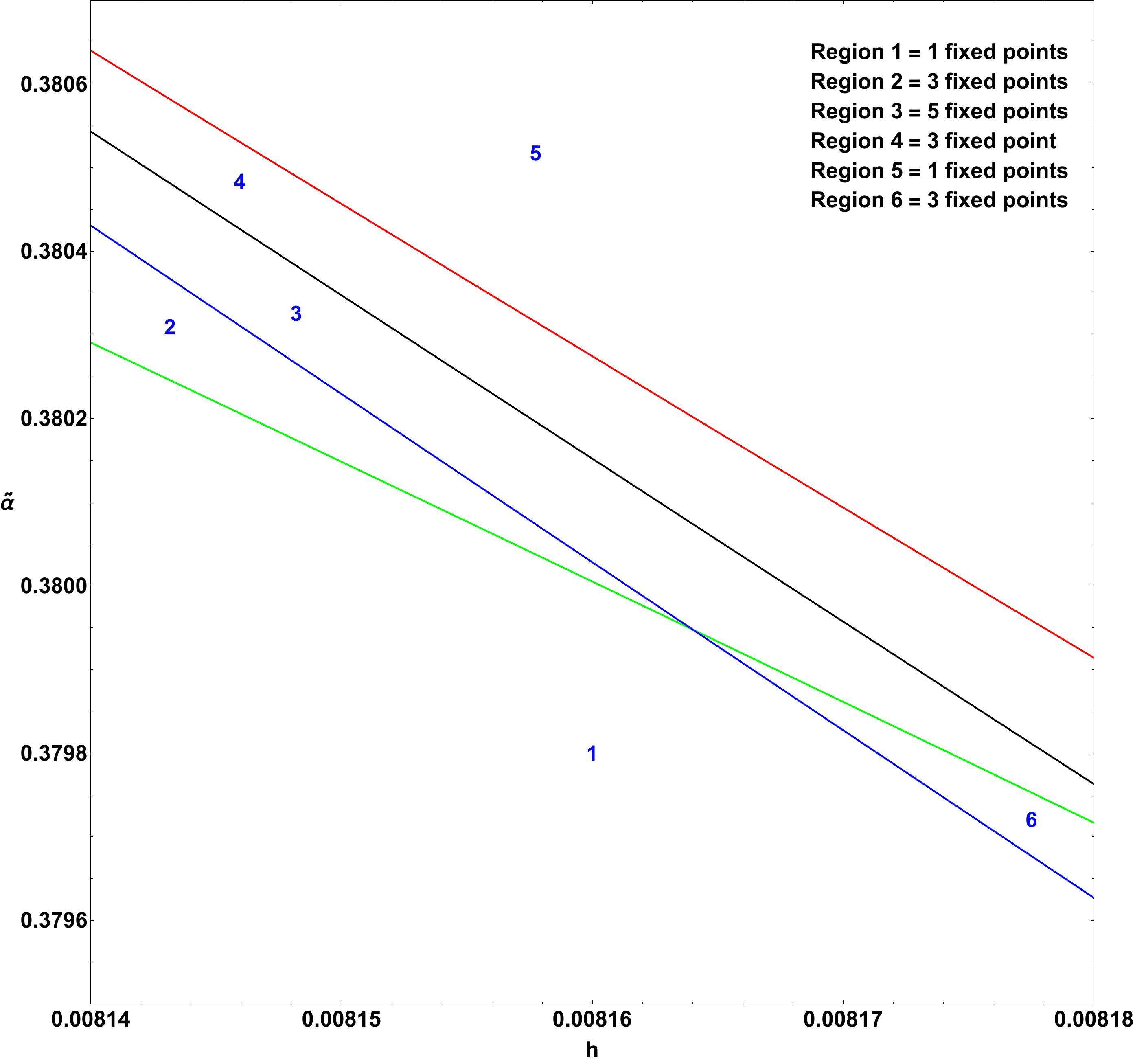}}
    \caption{Bifurcation diagrams of 6D GB black hole}
\end{figure*}

\section{Summary and Concluding Remarks}\label{secVI}

In this work, we have presented a perspective on black hole phase structure by formulating it within the framework of bifurcation theory which is a well-established approach in nonlinear dynamical systems. By introducing an appropriate bifurcating function, we have demonstrated how different classes of black hole exhibit characteristic bifurcation behavior, which enables a unified and systematic classification of their thermodynamic phases. The bifurcation diagrams reveal various types of fixed points whose stability can be directly linked to the thermodynamic stability of black hole's phases. Specifically, we have shown that stable fixed points correspond to thermodynamically stable black hole branches, whereas unstable fixed points indicate configurations prone to decay. Through numerical analysis of the bifurcation equation, we observed that stable black hole solutions exhibit vanishing slope over time, signifying convergence to a stable configuration. In contrast, dynamically unstable branches show divergence from equilibrium, indicating the natural decay of small black hole.In summary, our analysis demonstrates that black hole phase transition phenomena can be effectively understood through the study of bifurcations in the underlying dynamical system.\\

\begin{figure*}[t!]
\includegraphics[scale=0.4]{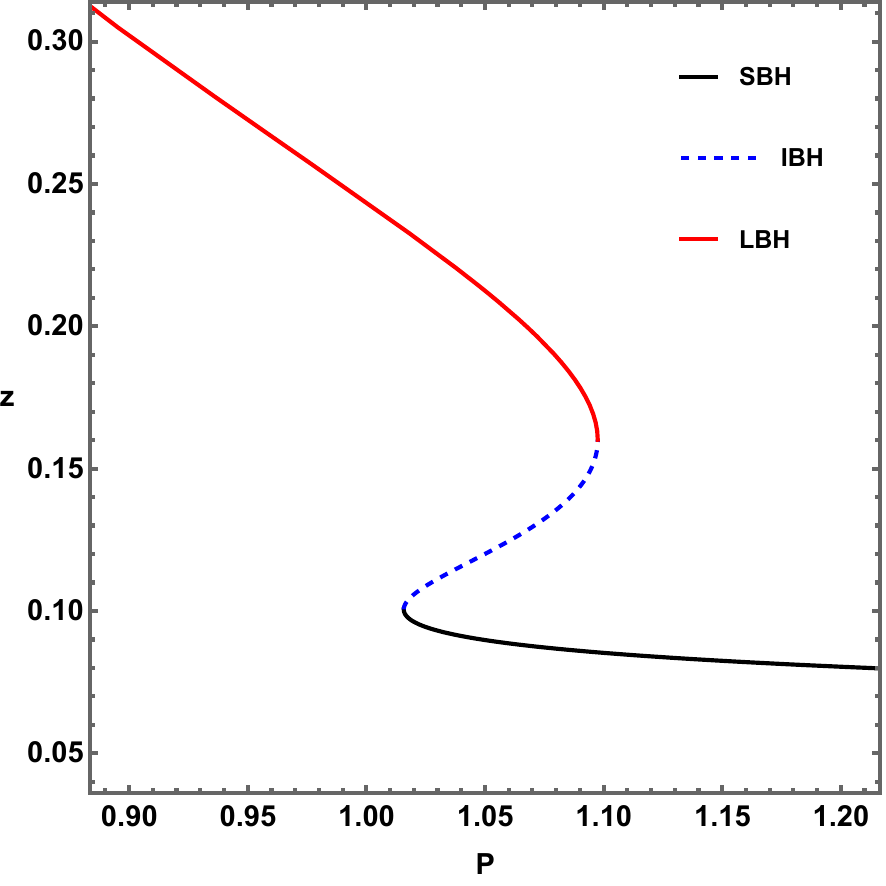}
\hspace{1cm}
\includegraphics[scale=0.47]{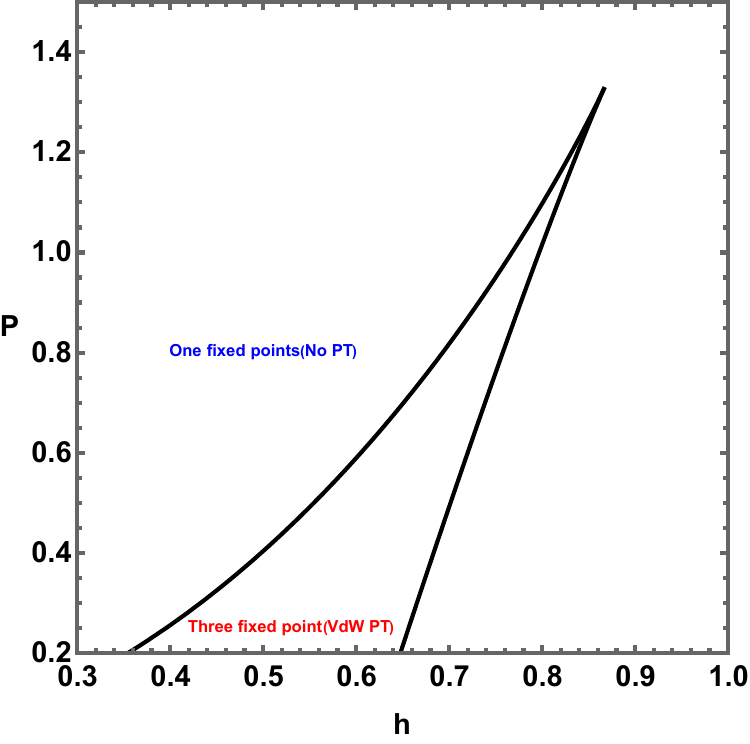}
\caption{Bifurcation diagram in extended phase space of RN AdS black hole}
\label{exte}
\end{figure*}
\begin{figure*}[t!]
\includegraphics[scale=0.425]{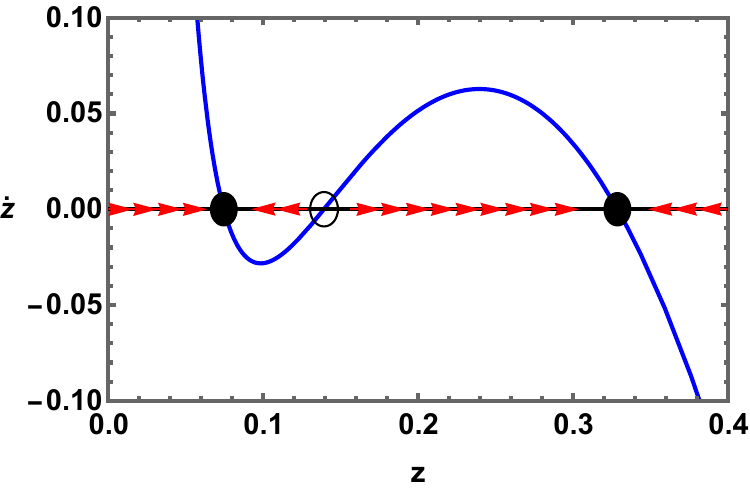}
\includegraphics[scale=0.42]{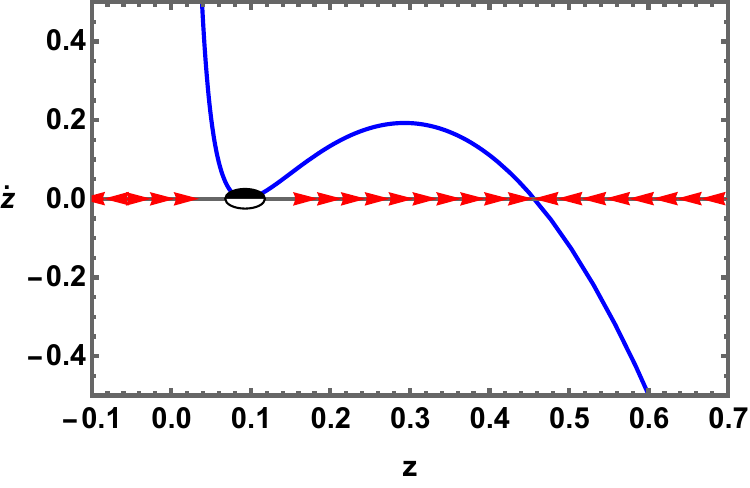}
\includegraphics[scale=0.42]{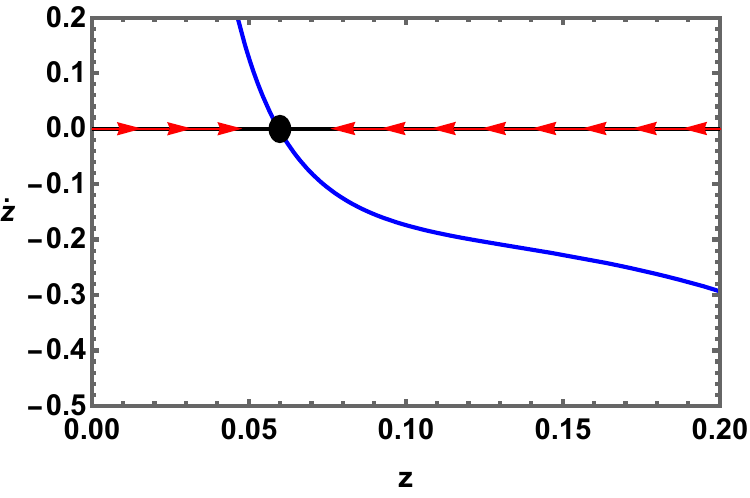}
\caption{Stability analysis of fixed point of RN AdS black hole in extended phase space}
\label{exte2}
\end{figure*}
One of the central motivations of this work is to demonstrate that bifurcation theory offers an alternative yet consistent framework for understanding black hole phase transitions.  In this framework, the bifurcation parameter $h$ plays a role analogous to the thermodynamic temperature, and variations in $h$ lead to qualitative changes in the number and stability of fixed points in the corresponding dynamical system.  Our analysis reveals that certain critical behaviors and phase transition types observed in black hole thermodynamics can naturally emerge from bifurcation theory. For instance, considering the extended phase space thermodynamics of the RN-AdS black hole, the bifurcation function can be defined as
\begin{equation}
\dot{z} = \frac{1}{2} \left(4 \pi h z - 8 \pi P z^2 + \frac{Q^2}{z^2} - 1\right),
\end{equation}
where $P$ represents the thermodynamic pressure.  By applying the  method outlined in the previous sections, the corresponding bifurcation diagrams are shown in Fig.~\ref{exte}, exhibit a close resemblance to the familiar $P$–$V$ criticality curves in extended black hole thermodynamics.  Specifically, the bifurcation curves between $z$ AND $P$ for the RN-AdS black hole display features that closely match the van der Waals-like phase transition structure.  Notably, the broken pitchfork bifurcation can  also be recovered in this scenario. The black holes exhibiting pitchfork-like bifurcation diagrams typically undergo first-order phase transitions analogous to those of a van der Waals fluid.  Furthermore, in cases where a cusp catastrophe structure arises,  this is indicative of second-order critical behavior. The $P$-$h$ diagram (see the right panel of Fig.\ref{exte}) illustrates regions in the parameter space where the number of fixed points varies. In particular, when three fixed points exist, the system exhibits a Van der Waals-type phase transition. Outside this region, no such transition occurs. The black solid line in the diagram marks the critical boundary, which matches well with the critical points obtained through conventional thermodynamic analysis.\\

\begin{figure*}[htb!]
\includegraphics[scale=0.425]{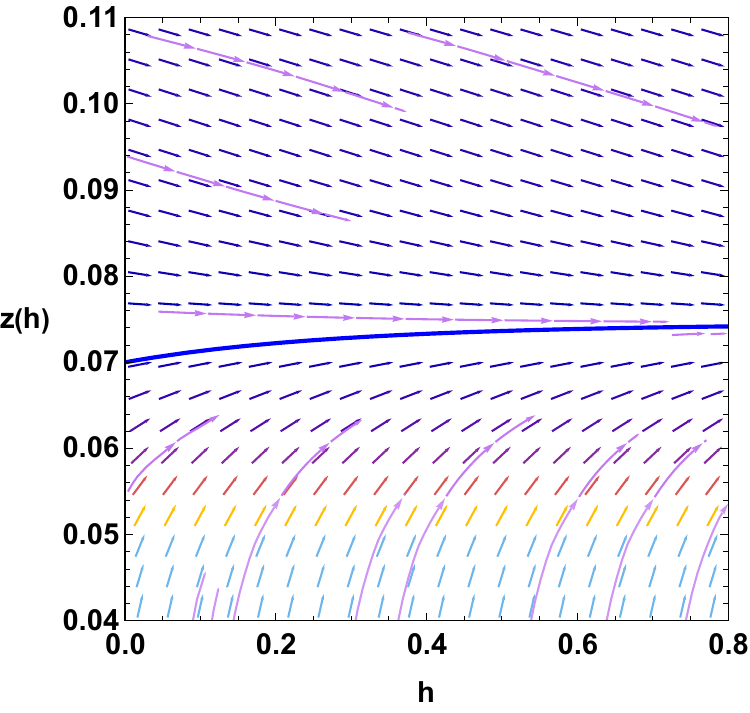}
\includegraphics[scale=0.42]{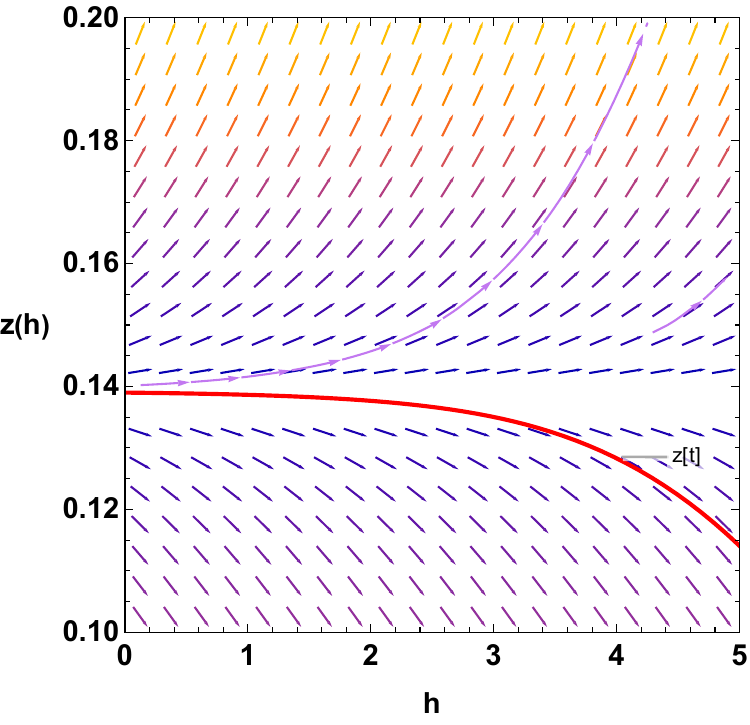}
\includegraphics[scale=0.42]{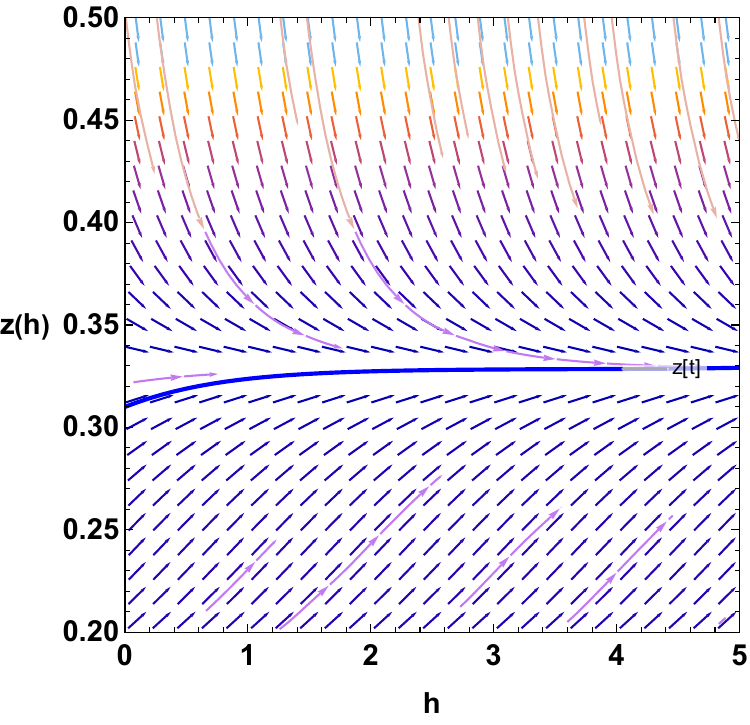}
\caption{Dynamical analysis of RN AdS black hole in extended phase space}
\label{exte3}
\end{figure*}

The stability of these fixed points is illustrated in Fig.\ref{exte2}. For parameter values $(P, h)$ lying inside the black curve in Fig.\ref{exte}, three fixed points emerge which corresponds to a stable,  an unstable, and another stable configuration,  that mimicks the characteristic behavior of RN AdS black holes undergoing a Van der Waals-type transition.  When the parameters $(P, h)$ lie on the black curve,  the system admits half stable or metastable fixed points. Finally, for parameter values outside the black boundary, the system admits a single stable fixed point, and no phase transition is observed. The dynamical stability of the black hole branches is presented in Fig.\ref{exte3}. This dynamical picture agrees well with the thermodynamic stability structure expected from the Van der Waals-like phase behavior of RN-AdS black holes. \\

Although the bifurcation diagrams are derived from a dynamical systems perspective, they successfully reproduce the well-established thermodynamic phase structures and critical phenomena in black hole systems. Additionally, we have shown that the thermodynamic stability of black hole branches corresponds directly to the dynamical stability of the associated fixed points. \\

The methodology we present is general and can be extended to a wide class of black hole spacetimes. For example, rotating solutions such as Kerr-AdS black holes also exhibit broken pitchfork bifurcations as shown in Fig.\ref{kerrads}, and the same systems, if studied under modified entropy frameworks (e.g., with Kaniadakis entropy) display higher-order bifurcations such as three-multifold saddle-node types. This method also can be applied for non AdS black hole also. For instance,  flat Kerr black hole shows two fixed points and one half stable point hence they can be classified under Saddle node bifurcation class as shown in Fig.\ref{kerr}. Our formalism can also be readily applied to higher-dimensional black holes, as demonstrated by the 6D Gauss-Bonnet-AdS case. We emphasize that most well-known black hole solutions can be systematically classified using the bifurcation schemes summarized in Table~1 of the manuscript. Furthermore, additional bifurcation types,  such as five-fold or six fold saddle node bifurcations,  may emerge naturally in black holes with nonlinear electrodynamic corrections or other modified gravity theories. Overall, the bifurcation framework presented here is both robust and flexible, and it provides a unified approach that can be applied to analyze phase structures across a broad spectrum of black hole models.

To make the classification scheme transparent and accessible, we have summarized the bifurcation structure in a tabular format in Table 1.  By following this scheme, black hole can be systematically classified into distinct bifurcation classes based on the number and stability of fixed points in the system.In future work, this bifurcation-based approach can be extended to more complex black hole scenarios.

\section*{Acknowledgements} 
BH would like to thank DST-INSPIRE, Ministry of Science and Technology fellowship program, Govt. of India for awarding the DST/INSPIRE Fellowship[IF220255] for financial support. 	

\begin{widetext}
\begin{figure*}[t!]
\centering{
    \subfloat[Bifurcation diagram : Kerr AdS black hole  We have kept h=0.3  fixed.\label{kerrads}]{
        \includegraphics[width=0.4\textwidth]{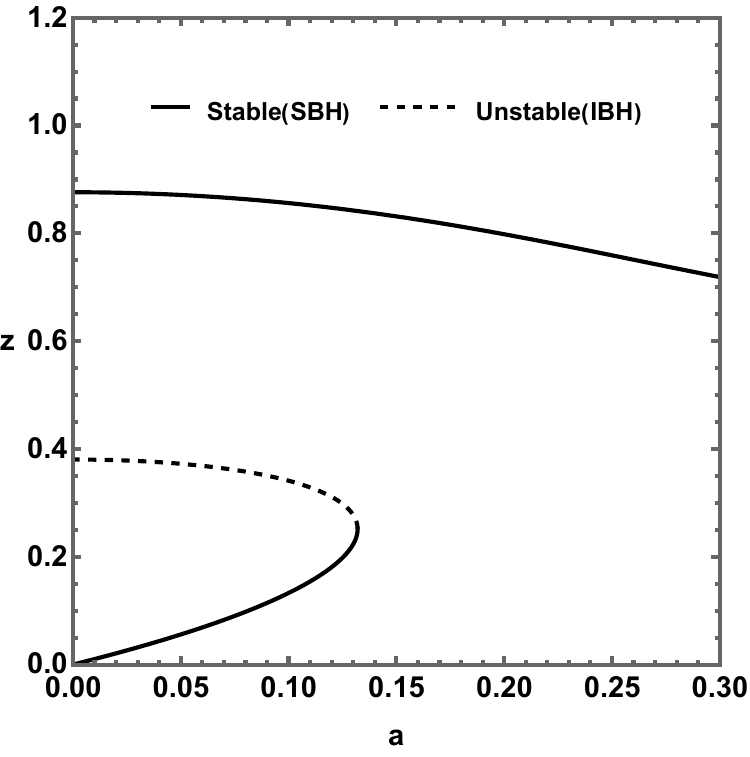}
    }
    \subfloat[Bifurcation diagram : Kerr black hole.  We have kept h=0.3  fixed.\label{kerr}]{
        \includegraphics[width=0.4\textwidth]{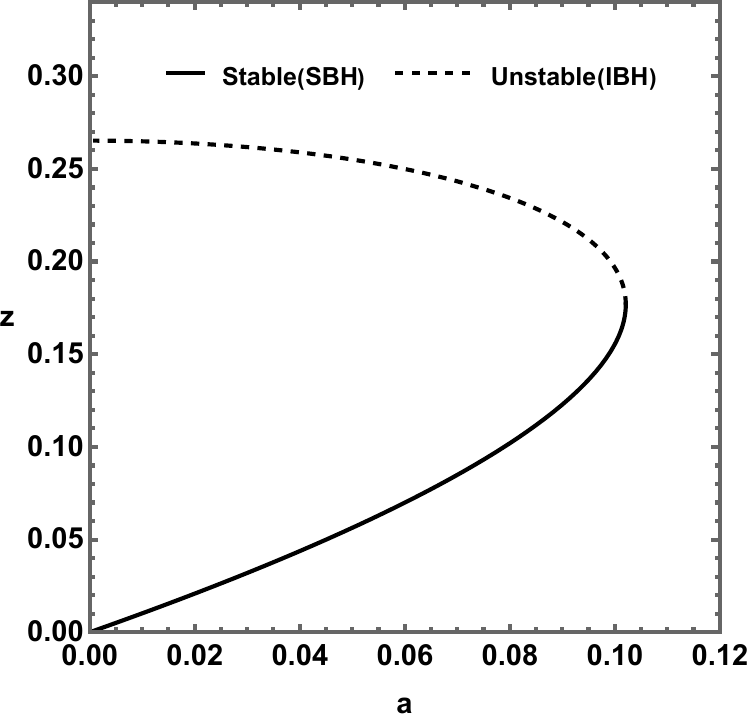}}
        }
    \caption{Bifurcation diagrams of Kerr black hole(both AdS and flat)}
\end{figure*}
\begin{table*}[t!]
\centering
\setlength{\tabcolsep}{6pt}
\begin{tabular}{|c|c|c|c|c|}
\hline
\textbf{Black Hole System} & \textbf{No. of Fixed Points} & \textbf{No. of Half-Stable Points} & \textbf{Bifurcation Type} \\
\hline
Schwarzschild-AdS & 2 & 1 & Saddle-node \\
\hline
RN AdS & 3 or 1 & 2 & Broken pitchfork \\
\hline
Euler Heisenberg AdS & 4, 2 or 0 & 3 & 3-Multifold saddle node \\
\hline
6D Gauss Bonnet AdS & 5, 3 or 1 & 4 & 4-Multifold saddle-node \\
\hline
\end{tabular}
\caption{Summary of bifurcation structures in various black hole systems studied. }
\label{tab1}
\end{table*}
\newpage
\end{widetext}

\end{document}